\documentclass[12pt]{article}
\usepackage{amsmath}
\usepackage{cite}
\usepackage{epsfig}
\usepackage{epstopdf}
\usepackage{amssymb}

\def\[{\left [}
\def\]{\right ]}
\def\({\left (}
\def\){\right )}

\def\r{\rho}

\def\r2{\sqrt{2}}

\newcommand{\bbibitem}[1]{\bibitem{#1}\marginpar{#1}}

\def\Label#1{\label{#1}%
  \smash{\hbox to0pt{\raise1ex\hbox{\tiny[#1]}\hss}}}
\def\noLabels{\let\Label=\label}
\def\nobbibitem{\let\bbibitem=\bibitem}

\newcommand{\bea}{\begin{eqnarray}}
\newcommand{\eea}{\end{eqnarray}}

\newcommand{\beq} {\begin{equation}}
\newcommand{\eeq} {\end{equation}}
\newcommand{\beqa} {\begin{eqnarray}}
\newcommand{\eeqa} {\end{eqnarray}}

\newcommand{\beqn}{\begin{eqnarray}}
\newcommand{\eeqn}{\end{eqnarray}}

\begin{document}

\begin{flushright}
HIP-2011-27/TH\\
Nordita-2011-90\\
RH-15-2011\\
UUITP-28/11
\end{flushright}

\vskip 2cm \centerline{\Large {\bf Thermalization and entanglement following}}
\vskip .5cm \centerline{\Large {\bf a non-relativistic holographic quench}}
\vskip 2cm
\renewcommand{\thefootnote}{\fnsymbol{footnote}}
\centerline
{{\bf Ville Ker\"{a}nen,$^{1,2}$
\footnote{vkeranen@nordita.org}, Esko Keski-Vakkuri,$^{3,4,5}$
\footnote{Esko.Keski-Vakkuri@helsinki.fi}, Larus Thorlacius,$^{1,2}$
\footnote{larus@nordita.org}
}}
\vskip .5cm
\centerline{\it
${}^{1}$Nordita}
\centerline{\it Roslagstullsbacken 23, SE-106 91 Stockholm, Sweden}
\centerline{\it
${}^{2}$ University of Iceland, Science Institute}
\centerline{\it Dunhaga 3, IS-107 Reykjavik, Iceland}
\centerline{\it
${}^{3}$Helsinki Institute of Physics }
\centerline{\it P.O.Box 64, FIN-00014 University of
Helsinki, Finland}
\centerline{\it
${}^{4}$Department of Physics}
\centerline{\it P.O.Box 64, FIN-00014 University of
Helsinki, Finland}
\centerline{\it
${}^{5}$Department of Physics and Astronomy, Uppsala University}
\centerline{\it SE-75108 Uppsala, Sweden}

\setcounter{footnote}{0}
\renewcommand{\thefootnote}{\arabic{footnote}}

\begin{abstract}

We develop a holographic model for thermalization following a quench
near a quantum critical point with non-trivial dynamical critical exponent.
The anti-de Sitter Vaidya null collapse geometry is generalized to
asymptotically Lifshitz spacetime. Non-local observables such as
two-point functions and entanglement entropy in this background then
provide information about the length and time scales relevant to thermalization.
The propagation of thermalization exhibits similar "horizon" behavior as has
been seen previously in the conformal case and we give a heuristic argument
for why it also appears here. Finally, analytic upper bounds are obtained for
the thermalization rates of the non-local observables.

\end{abstract}

\newpage
\section{Introduction}

The study of out of equilibrium processes in quantum field theory is an important
and interesting problem but the available theoretical toolbox in this area is limited,
especially when it comes to strongly interacting theories.
From the experimental point of view, cold atom systems provide a unique setting
for exploring the quantum dynamics of strongly correlated quantum
systems \cite{bloch}. As an example, it has become possible to set up a quantum
phase transition between a superfluid and a Mott insulator, with the help of optical
lattices \cite{greiner1}. One can even study the quantum dynamics of the system as the
height of the optical lattice is changed suddenly in time, leading to a quantum quench
from the superfluid phase to the Mott insulator phase \cite{greiner2}. In this manner
cold atom systems provide a way of experimentally realizing and simulating strongly correlated
quantum dynamics near quantum critical points.

Quenches near quantum critical points are particularly interesting from a theoretical
point of view, because the response of the system is expected to be universal and to
apply to many different physical systems. Indeed there has been a lot of interesting
theoretical work studying quench dynamics near quantum critical
points \cite{Calabrese:2005in,Calabrese:2006,Calabrese:2007rg,DeGrandi} (for a review
see \cite{jacek}).
The work in \cite{Calabrese:2005in,Calabrese:2006} provided a
nice heuristic picture of the dynamics of two point correlation functions and
entanglement entropy after a quantum quench in 1+1 dimensional conformal field
theory (CFT). The heuristic picture, which we will refer to as the "horizon" effect,
starts with the production of a large number of "quasiparticles" from the quench,
as the ground state of the initial Hamiltonian is a highly excited state of the
quenched Hamiltonian. Quasiparticles that originate from adjacent points are
entangled through the initial ground state. As they travel at the speed of light after
the quench, they lead to correlations in non-local observables, that propagate on
a light cone. Inside the light cone, the observables average to their thermal values
while causality forces correlations outside the light cone to take the same form as
in the initial vacuum.
In \cite{Calabrese:2006} it was found that for free lattice systems the "horizon"
was smoothed out in the sense that, while thermalization started at the group velocity
of the fastest quasiparticle, complete thermalization was only achieved at
the group velocity of the slowest quasiparticle.

Holography provides a novel way to study strongly correlated quantum systems
and has been used in \cite{AbajoArrastia:2010yt,Albash:2010mv,Balasubramanian:2010ce,Balasubramanian:2011ur,Aparicio:2011zy,Balasubramanian:2011at,Allais:2011ys} to study quenches in strongly coupled CFT's in
1+1 as well as 2+1 and 3+1 dimensions. A similar "horizon" effect has been observed
within the holographic approach. Alternative holographic models of quenches have been constructed
and studied in \cite{Das:2010yw,Basu:2011ft}.

The powerful methods of CFT that have been used to study quantum quenches
apply to "relativistic" theories with an underlying conformal symmetry. Generic
quantum critical points in condensed matter systems are, however, not conformally
invariant but instead exhibit anisotropic scaling of the form,
\beq
(\textbf{x},t)\rightarrow (\lambda\textbf{x},\lambda^z t),
\eeq
with a non-trivial dynamical critical exponent $z>1$. To study holographic duals of
these more general quantum critical points it was suggested \cite{Kachru:2008yh} to
consider gravitational theories in a spacetime that asymptotes to the
form\footnote{See also \cite{Koroteev:2007yp} for early work on gravitational
backgrounds with anisotropic scaling.}
\beq
ds^2=-r^{2z}dt^2+r^2d\textbf{x}^2+\frac{dr^2}{r^2},\label{eq:2}
\eeq
where the scaling symmetry $(\textbf{x},t)\rightarrow (\lambda\textbf{x},\lambda^z t)$
is realized as an isometry of the metric (\ref{eq:2}), when combined with the scaling
$r\rightarrow\lambda^{-1}r$. As is customary, we will refer to a metric that asymptotes
to (\ref{eq:2}) as an asymptotically Lifshitz metric and the aim of this paper is to study
quantum quenches for $z>1$ using holography in asymptotically Lifshitz spacetimes.
A priori it is not at all clear whether one should see a similar picture arise as in the case
of $z=1$. In particular, the quasiparticle dispersion relation (assuming the concept of a
quasiparticle makes sense in these theories), $\omega\propto k^z$, suggests that
thermalization could start infinitely fast but take an infinite time to complete. We will
see, however, that this is not what happens in the holographic model we consider here.
Instead, interactions change the picture qualitatively, and we recover a "horizon" effect
with a characteristic velocity $v\propto T^{(z-1)/z}$.

A further motivation for studying quenches in asymptotically Lifshitz spacetimes is that
some holographic models used to study field theories at finite density, give rise to
geometries that approach the Lifshitz spacetime (\ref{eq:2}) in the infrared limit of
small $r$ \cite{Hartnoll:2011fn}. Examples of such IR Lifshitz finite density systems
are holographic superfluids in their ground state \cite{Gubser:2009cg} and
"electron stars"  \cite{Hartnoll:2009ns,Hartnoll:2010gu}. Thus our results may shed
some light on the non-equilibrium dynamics of holographic finite density systems.

A quench can be achieved by suddenly changing the values of coupling constants in
a given field theory. In holography, the coupling constants of the dual field theory are
related to asymptotic values of bulk fields as $r\rightarrow\infty$. A sudden change in
the values of coupling constants leads to translationally invariant shifts in the fields
near $r\rightarrow\infty$. Since an asymptotically Lifshitz spacetime acts as a
gravitational potential well, in a manner analogous to AdS spacetime, the field
excitations are pulled towards $r=0$ and quickly accelerate to a speed very close to
that of light. Soon after the quench, one is left with a sharp shell of energy density
starting from $r\rightarrow \infty$ and falling into the bulk at the speed of light.
Eventually the shell of energy will form a black hole in the bulk, which corresponds
to approaching thermal equilibrium in the dual field theory. (There are also alternative
ways to achieve a quench holographically, see \cite{Das:2010yw,Basu:2011ft}.)

A simple model for a shell falling at the speed of light is given by the Vaidya
spacetime \cite{Vaidya:1951zz,Bonnor:1970zz}. Indeed the work in \cite{Bhattacharyya:2009uu} showed that for sufficiently small amplitude quenches of marginal operators, the spacetime is well approximated by the AdS-Vaidya metric. The asymptotically Anti-de Sitter
Vaidya geometry was used to study holographic quenches in \cite{AbajoArrastia:2010yt,Balasubramanian:2010ce,Albash:2010mv,Balasubramanian:2011ur,Balasubramanian:2011at,Allais:2011ys,Ebrahim:2010ra}. In this paper we generalize the Vaidya metric to an asymptotically
Lifshitz spacetime. We do this in the context of the family of models studied in
\cite{Taylor:2008tg,Tarrio:2011de}, which have Einstein's gravity, a dilaton and some
number of U(1) gauge fields in the bulk. We then consider non-local probes (correlation
functions of gauge invariant operators and holographic entanglement entropy) in the
quench geometry to uncover the characteristic time and length scales of the thermalization
process. More precisely, we will analyze lengths of geodesics and minimal surface areas and assume that the usual holographic proposal
\cite{Ryu:2006bv,Ryu:2006ef,Hubeny:2007xt} extends from asymptotically AdS spacetime
to Lifshitz scaling so that we can interpret the results
as entanglement entropy in the dual field theory. Our results may also be relevant for work
studying entanglement entropy in non-relativistic field theory, see \cite{Solodukhin:2009sk,Nesterov:2010yi,deBoer:2011wk}.

The outline of the paper is as follows. In Section~2, we briefly review the black hole
solutions found in \cite{Tarrio:2011de} and in Section~3, we construct Vaidya type
solutions that describe a holographic quench.  In Section~4, we study equal time two
point functions in the quench state, using the geodesic approximation. We also derive
an analytic upper bound for the velocity at which the thermalization of the two point
function spreads. In Section~5, we study the entanglement entropy in the quench state
and again derive an upper bound for the velocity at which the thermalization of the entanglement entropy can spread.

\newpage

\section{Static black holes}

We will consider the gravitational theory specified by the action
\beq
S= \frac{1}{16\pi G_N}\int d^4x\sqrt{-g}\Bigg[R-2\Lambda-\frac{1}{2}(\partial\phi)^2-\frac{1}{4}\sum_{i=1}^Ne^{\lambda_i\phi}F_i^2\Bigg].
\eeq
This theory has a family of Lifshitz solutions. First we will consider the case when the number of $U(1)$ gauge fields in the bulk is $N=2$. The generalization to a larger number of spacetime dimensions and to larger number of $U(1)$ gauge fields is straightforward. Throughout we will denote Einstein's equations and Maxwell's equations as
\beq
E_{\mu\nu}=R_{\mu\nu}-\frac{1}{2}R g_{\mu\nu}+\Lambda g_{\mu\nu}-T_{\mu\nu}=0,\quad M^{\nu}_i=D_{\mu}(e^{\lambda_i\phi}F^{\mu\nu}_i)=0.\label{eq:eqs}
\eeq
First we will review the static Lifshitz solutions obtained in \cite{Tarrio:2011de}. The three parameter family of solution found in \cite{Tarrio:2011de} is
\begin{align}
ds^2&=-r^{2z}b(r)dt^2+\frac{dr^2}{r^2 b(r)}+r^2d\textbf{x}^2,\nonumber
\\
b(r)&=1-mr^{-(2+z)}+\frac{\rho_2^2 \mu^{-\sqrt{(z-1)}}}{4z}r^{-2(1+z)},\nonumber
\\
A^{(1)}_t&=\sqrt{\frac{2(z-1)}{2+z}}\mu^{1/\sqrt{z-1}}r^{2+z},\label{eq:3}
\\
A^{(2)}_t&=-\frac{\rho_2\mu^{-\sqrt{z-1}}}{z}r^{-z},\nonumber
\\
e^{\phi}&=\mu r^{2\sqrt{z-1}},\nonumber
\end{align}
specified by the parameters $(\mu,\rho_2,m)$. For (\ref{eq:3}) to be a solution to the equations of motion, the value of the cosmological constant and the parameters $\lambda_i$ must be fixed as
\beq
\Lambda=-\frac{(2+z)(1+z)}{2},\quad \lambda_1=-\frac{2}{\sqrt{z-1}},\quad \lambda_2=\sqrt{z-1}.
\eeq
The case $\rho_2=m=0$ corresponds to the Lifshitz vacuum solution. Before constructing the infalling shell solutions, it is convenient to write the solution (\ref{eq:3}) in an Eddington-Finkelstein-like (EF) coordinate system.

The ingoing null geodesic is easily found from the metric in (\ref{eq:3}) as
\beq
dt+\frac{r^{-z-1}}{b(r)}dr=0.\label{eq:nullgeod}
\eeq
To define the EF coordinate system, we define a new time coordinate through the relation
\beq
dv=dt+\frac{r^{-z-1}}{b(r)}dr.\label{eq:ef}
\eeq
Next we exchange the $t$ coordinate with the new $v$ coordinate to obtain the metric
\beq
ds^2=-r^{2z}b(r)dv^2+2 dv dr r^{z-1}+r^2d\textbf{x}^2.
\eeq
The gauge fields in the solution (\ref{eq:3}) are of the form $A^{(i)}=A_t^{(i)}(r)dt$. When going to the new $v$ coordinate system these become
\beq
A^{(i)}=A^{(i)}_t(r)(dv- \frac{r^{-z-1}}{b(r)}dr) \ ,
\eeq
so that there is an $A_r$ component induced in the EF coordinate system. For future convenience we note that the $A_r$ component can be set to vanish with a gauge transformation. Thus, the gauge fields in the new coordinate system have the form $A=A^{(i)}_t dv$, where $A^{(i)}_t$ are the fields specified in (\ref{eq:3}).

\newpage
\section{The infalling shell solutions}

An infalling shell of massless and pressureless charged matter in an asymptotically flat spacetime is described by the Vaidya metric \cite{Vaidya:1951zz,Bonnor:1970zz}, which is known analytically. The Vaidya metric corresponds to a Reissner-Nordstr\"{o}m black hole, with the free parameters corresponding to the mass and the charge promoted into functions of the EF time coordinate $v$. The Vaidya spacetime is sourced by an explicit energy momentum tensor for massless null matter $T_{vv}\neq 0$ and a current density $j_{v}\propto j^r\neq 0$. Our task is to find a similar solution as the Vaidya solution, which asymptotes to an asymptotically Lifshitz spacetime. Again we will have an explicit source in the $vv$ component of Einstein's equations and in the $r$ component of Maxwell's equations, while all the other Einstein's equations as well as the matter equations will be solved without explicit sources.

Motivated by the conventional Vaidya solution \cite{Bonnor:1970zz} we choose the following ansatz for the metric and the matter fields
\begin{align}
ds^2&=-r^{2z}b(r,v)dv^2+2r^{z-1}dvdr+r^2d\textbf{x}^2,\nonumber
\\
b(r,v)&=1-m(v)r^{-(z+2)}+f(v)\frac{\rho_2^2 \mu^{-\sqrt{(z-1)}}}{4z}r^{-2(1+z)},\nonumber
\\
A^{(1)}_v&=h_1(v)\sqrt{\frac{2(z-1)}{2+z}}\mu^{1/\sqrt{z-1}}r^{2+z},\label{eq:4}
\\
A^{(2)}_v&=-h_2(v)\frac{\rho_2\mu^{-\sqrt{z-1}}}{z}r^{-z},\nonumber
\\
e^{\phi}&=h_3(v)\mu r^{2\sqrt{z-1}},\nonumber
\end{align}
where $m(v),f(v),h_1(v),h_2(v),h_3(v)$ are arbitrary functions of $v$ for the moment. Substituting the ansatz (\ref{eq:4}) into Einstein's equations leads to relations between the above functions of $v$. The $vr$ component of the Einstein's equations leads to
\beq
h_1=h_3^{1/\sqrt{z-1}},\quad f=h_2^2h_3^{\sqrt{z-1}}.
\eeq
The $xx$ and $yy$ components of Einstein's equations are
\beq
E_{xx}=E_{yy}=\sqrt{z-1}r^{2-z}\frac{h_3'(v)}{h_3(v)}.
\eeq
In order to solve them (for $z\neq1$), we must set
\beq
h_3(v)=const.
\eeq
Furthermore we can without loss of generality set $h_3=1$. This also leads to $h_1=1$ and
\beq
f=h_2^2.
\eeq
We are thus left with two arbitrary functions $m(v)$ and $h_2(v)$. Since $h_2(v)$ appears only in the combination $h_2(v)\rho_2$, we see that our solution is simply equivalent to promoting $m$ and $\rho_2$ into arbitrary functions of $v$
\begin{align}
ds^2&=-r^{2z}b(r,v)dv^2+2r^{z-1}dvdr+r^2d\textbf{x}^2,\nonumber
\\
b(r,v)&=1-m(v)r^{-(z+2)}+\frac{\rho_2(v)^2 \mu^{-\sqrt{(z-1)}}}{4z}r^{-2(z+1)},\nonumber
\\
A^{(1)}_v&=\sqrt{\frac{2(z-1)}{2+z}}\mu^{1/\sqrt{z-1}}r^{2+z},\label{eq:5}
\\
A^{(2)}_v&=-\frac{\rho_2(v)\mu^{-\sqrt{z-1}}}{z}r^{-z},\nonumber
\\
e^{\phi}&=\mu r^{2\sqrt{z-1}},\nonumber
\end{align}
We will refer to (\ref{eq:5}) as the Lifshitz-Vaidya solution. The fields in (\ref{eq:5}) do not yet solve all of the Einstein's and Maxwell's equations, but there are the non-vanishing components
\begin{align}
E_{vv}&=-\frac{\mu^{-\sqrt{z-1}}r^{-2-z}}{2z}\rho_2(v)\rho_2'(v)+\frac{m'(v)}{r^2},\label{eq:eins}
\\
M^{r}_2&=r^{-1-z}\rho_2'(v).\label{eq:maxws}
\end{align}
These non-vanishing components can be identified as sources from charged infalling massless matter with a vanishing pressure. The energy momentum tensor of such matter has the form $T_{\mu\nu}=\rho u_{\mu}u_{\nu}$ and a current density $J_{\mu}=\rho_e u_{\mu}$ with $u_{\mu}=\delta_{\mu v}$. Indeed this is of the form we need to solve the Einstein's and Maxwell's equations (\ref{eq:eins}) and (\ref{eq:maxws}) with
\begin{align}
\rho&=-\frac{\mu^{-\sqrt{z-1}}r^{-2-z}}{2z}\rho_2(v)\rho_2'(v)+\frac{m'(v)}{r^2},
\\
\rho_e&=\frac{\rho_2'(v)}{r^2}.
\end{align}

We have a solution for two $U(1)$ gauge fields, but it is trivial to reduce it to the case of a single gauge field studied in \cite{Taylor:2008tg} by setting $\rho_2=0$, which makes the $A^{(2)}$ field vanish identically. We can also find solutions for theories with more $U(1)$ gauge fields in \cite{Tarrio:2011de} by simply promoting the corresponding free charge densities $\rho_i$, for $i>1$, and the energy density $m$ into functions of $v$. Another direction of generalization is to consider different values of the spacetime dimension. This generalization seems also to work trivially and we have confirmed this for the bulk spacetime dimensions $d=4,5,6$. The explicit solution in the general case is simply \cite{Tarrio:2011de}
\begin{align}
ds^2&=-r^{2z}b(r,v)dv^2+2r^{z-1}dvdr+r^2d\textbf{x}^2,\nonumber
\\
b(r,v)&=1-m(v)r^{-(z+d-2)}+\sum_{j=2}^{N}\frac{\rho_j(v)^2 \mu^{-\sqrt{2\frac{z-1}{d-2}}}}{2(d-2)(d+z-4)}r^{-2(d+z-3)},\nonumber
\\
\partial_rA^{(1)}_v&=\sqrt{2(d+z-2)(z-1)}\mu^{\sqrt{\frac{d-2}{2(z-1)}}}r^{d+z-3},\label{eq:5}
\\
\partial_rA^{(j)}_v&=\rho_j(v)\mu^{-\sqrt{2\frac{z-1}{d-2}}}r^{3-d-z},\quad(j=2,...,N)\nonumber
\\
e^{\phi}&=\mu r^{\sqrt{2(d-2)(z-1)}},\nonumber
\end{align}
where now $d=4,5,6$. This corresponds to the choice of parameters
\beq
\Lambda=-\frac{1}{2}(d+z-2)(d+z-3),\quad\lambda_1=-\sqrt{2\frac{d-2}{z-1}},\quad \lambda_j=\sqrt{2\frac{z-1}{d-2}},
\eeq
where $j=2,...,N$. It is straightforward to show that such an ansatz indeed solves the equations of motion.

\newpage
\section{2-point correlation functions}

In this section we study 2-point correlation functions in the quench state. For simplicity, we restrict our attention in what follows to uncharged black brane solutions with $A^{(2)}=0$
and the metric function
\beq
b(r,v)=1-m(v)r^{-d-z+2}.
\eeq
This allows us to focus on the differences that arise between our $z>1$ solutions and the previously studied $z=1$ case without the added complication of non-vanishing gauge charge. We expect our calculations to carry over to the charged case in a straightforward way. Furthermore we will focus on the case $d=4$. Higher dimensional cases are discussed in Appendix A, where as an example we show the results for $d=5$ and $z=2$.

For the convenience of numerical computations we define a new radial coordinate $u=1/r$
(not to be confused with an
Eddington-Finkelstein coordinate), in terms of which the metric of interest reads
\beq
ds^2=-u^{-2z}b(u,v)dv^2-2u^{-1-z}du dv+u^{-2}d\textbf{x}^2,\quad b(u,v)=1-m(v)u^{2+z}.
\eeq

The calculation of the correlation function is performed in the geodesic approximation, which becomes
more accurate as the scaling dimensions of the corresponding operators in the correlation function are increased. The idea of the geodesic approximation is to perform a saddle point approximation in the path integral over particle paths in the bulk \cite{Balasubramanian:1999zv}, to obtain the bulk Feynman propagator. As was discussed in \cite{Banks:1998dd} one can obtain the boundary theory correlator by pulling the points in the bulk Feynman propagator to the boundary and multiplying by appropriate powers of the cutoff $\epsilon$. This leads to the boundary theory two point function in the geodesic approximation as given by
\beq
\langle\mathcal{O}(x)\mathcal{O}(x')\rangle \approx \epsilon^{-2\Delta} e^{-\Delta\int d\tau\sqrt{g_{\mu\nu}\frac{dx^{\mu}}{d\tau}\frac{dx^{\nu}}{d\tau}}},\label{eq:geodcor}
\eeq
where $x^{\mu}(\tau)$ is the geodesic with minimal length and $\Delta$ is the mass of the bulk particle, or equivalently the scaling dimension of the dual operator. There is no factor of $i$ in the exponent in (\ref{eq:geodcor}) because we are specializing to spacelike geodesics, and it is convenient
to factor out a minus sign from the square root.  By symmetry we will choose the geodesics to have $y=$const, where $\textbf{x}=(x,y)$ are coordinates in the transverse plane, and furthermore parametrize it with the coordinate $x$ so that
\beq
u=u(x),\quad v=v(x).
\eeq
With this ansatz the particle action becomes
\beq
S=\Delta\int dx L=\Delta\int dx\sqrt{u^{-2}-u^{-2z}b(u,v)(v'(x))^2-2 u^{-z-1}v'(x)u'(x)}.\label{eq:length}
\eeq
Since the Lagrangian $L$ does not depend explicitly on $x$ there is a conserved Hamiltonian
\beq
H=\frac{\partial L}{\partial u'}u'+\frac{\partial L}{\partial v'}v'-L=-\frac{1}{u^2 L}.\label{eq:hamilton}
\eeq
To find the geodesics we need to set up boundary conditions. In this work we will compute only equal time correlation functions\footnote{For correlation functions with unequal times in the case of a holographic quench for $z=1$ see \cite{Aparicio:2011zy}.}. The geodesic will start from $x=-l/2$ and end at $x=l/2$. Furthermore there is a turning point at $x=0$ around which the minimal length geodesic is symmetric. Regularity imposes the following boundary conditions at the turning point
\beq
u'(0)=0=v'(0).
\eeq
We will furthermore denote the turning point as $u(0)=u_*$ and $v(0)=v_*$. With the above boundary conditions we can calculate the value of the Hamiltonian
\beq
H=-u_*^{-1}.
\eeq
As the set of two independent equations of motion we will use one of the Euler-Lagrange equations and the conserved Hamiltonian (\ref{eq:hamilton})
\begin{align}
uv''+2v'u'-u^{z-1}-\frac{1}{2}u^{z+2}\partial_u(u^{-2z}b(u,v))(v')^2&=0,\label{eq:geo1}
\\
1-2u^{1-z}u'v'-u^{2-2z}b(u,v)(v')^2-\frac{u_*^2}{u^2}&=0.\label{eq:geo2}
\end{align}
For the purpose of performing numerics it is convenient to transform to dimensionless variables as follows
\beq
\tilde{u}=r_0 u,\quad \tilde{v}=r_0^z v,\quad \tilde{x}=r_0 x,\label{eq:scaling}
\eeq
where $r_0$ is defined in such a way that the position of the horizon of the forming black hole is at $u=1$. In this way we can relate $r_0$ to the temperature at late time thermal equilibrium as
\beq
r_0=\Big(\frac{4\pi T}{2+z}\Big)^{1/z}.
\eeq
In what follows we will work with dimensionless coordinates and drop the tildes. After the change of variables to dimensionless coordinates, the equations for the geodesics (\ref{eq:geo1}) and (\ref{eq:geo2}) stay invariant in their form except that now
\beq
b(v,u)=1-u^{2+z}\frac{m(v/r_0^z)}{m(\infty)}.
\eeq

\subsection{Lifshitz vacuum}

To begin with, we calculate the equal time correlator in the zero temperature Lifshitz spacetime, i.e. we set $b=1$. Due to time translational invariance, the geodesic has to be independent of time $t$ since "momentum" conservation in the $t$ direction forbids the geodesic to turn around in time, which again is necessary to satisfy the boundary condition of having both ends of the geodesic at the same time. Thus, the geodesic has to be independent of time $t$ and we can use $dv=dt-u^{z-1}du=-u^{z-1}du$. Substituting this into (\ref{eq:geo1}) leads to
\beq
\frac{d}{dx}(uu')=-1,
\eeq
which can be easily integrated into
\beq
u=\sqrt{D+2C x-x^2}.
\eeq
Furthermore by imposing the boundary conditions $u'(0)=0$ and $u(l/2)=0$ we get
\beq
u(x)=\sqrt{\frac{l^2}{4}-x^2}.\label{eq:vacuumgeo}
\eeq
Using the conserved Hamiltonian we can write the on shell action simply as
\beq
S=\Delta\int dx\frac{u_*}{u(x)^2}.\label{eq:action2}
\eeq
Substituting (\ref{eq:vacuumgeo}) into (\ref{eq:action2}) leads to
\beq
S=\Delta l\int_{0}^{l/2}\frac{dx}{l^2/4-x^2}.
\eeq
This integral is divergent near $x=l/2$ and must be regulated. We regulate it by introducing a cutoff for $x$ at $l/2-\tilde{\epsilon}$. In this way the on shell action becomes
\beq
S=\Delta l\int_{0}^{l/2-\tilde{\epsilon}}\frac{dx}{l^2/4-x^2}=2\Delta \tanh^{-1}(1-\frac{2\tilde{\epsilon}}{l})\approx -\Delta\log\Big(\frac{\tilde{\epsilon}}{l}\Big).
\eeq
The cutoff in $x$ can be related to a cutoff $\epsilon$ in the holographic coordinate $u$ through $u(l/2-\tilde{\epsilon})=\epsilon$, which leads to the relation $\epsilon=\sqrt{\tilde{\epsilon} l}$. In this way the two point correlator becomes
\beq
G_2(l,t)=\langle\mathcal{O}(-l/2,t)\mathcal{O}(l/2,t)\rangle\approx \epsilon^{-2\Delta}e^{-S}=\frac{1}{l^{2\Delta}}\label{eq:vaccorrelator},
\eeq
which is seen to be independent of the Lifshitz scaling exponent $z$. This follows simply because the spatial part of the Lifshitz spacetime metric is independent of $z$.

\subsection{Thermal equilibrium}

\begin{figure}[h]
\begin{center}
\includegraphics[scale=1.2]{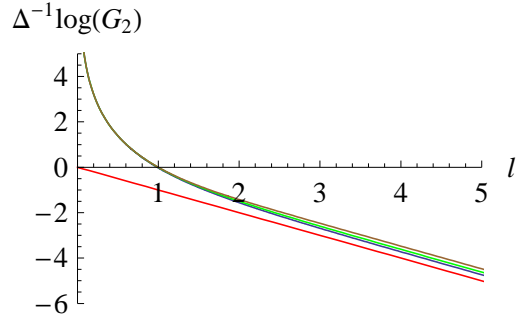}
\caption{\label{fig:thermalcorrelator} Logarithm of the thermal correlator for different values of z. The different curves correspond to $z=1,2,6$, from bottom to up. The red line is a line with slope $-1$. The figure shows that the thermal correlation function is fairly independent of the value of $z$.}
\end{center}
\end{figure}

The finite temperature equal time two point function can be obtained by solving (\ref{eq:geo1}) and (\ref{eq:geo2}) with the black hole metric factor
\beq
b(u)=1- u^{2+z}.
\eeq
We can solve (\ref{eq:geo1}) and (\ref{eq:geo2}) numerically with this choice for $b$. The result is shown in Fig. \ref{fig:thermalcorrelator}.
The main point is that the correlator behaves as
\beq
G_2(l,t)\propto e^{- l/\xi},\label{eq:thermalcorrelator}
\eeq
for sufficiently large $l$. The coefficient $\xi$ is the thermal correlation length, which is related to the temperature as
\beq
\xi= \Delta^{-1}\Big(\frac{4\pi T}{2+z}\Big)^{-1/z}.
\eeq.

\subsection{The quench}

Next we can solve (\ref{eq:geo1}) and (\ref{eq:geo2}) for the time dependent Lifshitz-Vaidya background with
\beq
b(u,v)=1-m(v)u^{2+z},
\eeq
where we choose the profile
\beq
m(v)=\frac{1}{2}(1-\tanh(v/v_0)).\label{eq:mass}
\eeq
For the time scale $v_0$ appearing in (\ref{eq:mass}) we will choose the value $v_0=10^{-2}$, or in dimensionful coordinates $v_0=10^{-2}(2+z)/(4\pi T)$. In the dual field theory this corresponds to adding a pulse of energy to the vacuum at the time $t=0$. The time scale in which the energy pulse appears is $v_0$. We call this operation a quench. Before solving the equations numerically we can understand some of the main features without the detailed calculation. At times earlier than $t=0$, the spacetime will look like the pure Lifshitz vacuum and thus, the equal time correlator will take the vacuum form (\ref{eq:vaccorrelator}).
Also for times much larger than $t=0$ the spacetime looks like the Lifshitz black hole so that one might think that the equal time correlator will take the thermal equilibrium form (\ref{eq:thermalcorrelator}). This conclusion does not generally hold for late times. Instead the
qualitative behavior of the correlator depends on the transverse separation $l$. If $l$ is
sufficiently small, the geodesic will not "drop" too close to the horizon and $v(x)>0$ along
the entire geodesic. Then the geodesic will indeed be that of the late-time black hole spacetime and one recovers the thermal correlator (\ref{eq:thermalcorrelator}).
On the other hand, when $l$ is sufficiently large, the geodesic passes through both the
event horizon and the apparent horizon \cite{AbajoArrastia:2010yt}, as discussed below, and part of the geodesic
will have $v(x)<0$. This means that the geodesic passes through the infalling shell at
$v=0$ and extends into the part of the spacetime with vacuum geometry. In this case the correlator will not be thermal.

\begin{figure}[h]
\begin{center}
\includegraphics[scale=1.0]{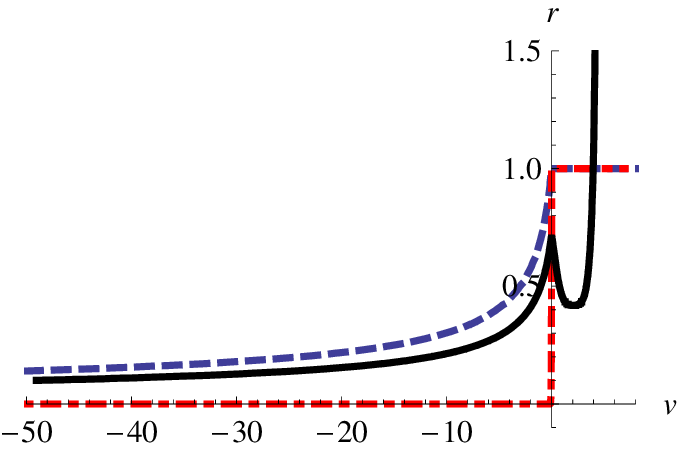}
\quad
\includegraphics[scale=1.0]{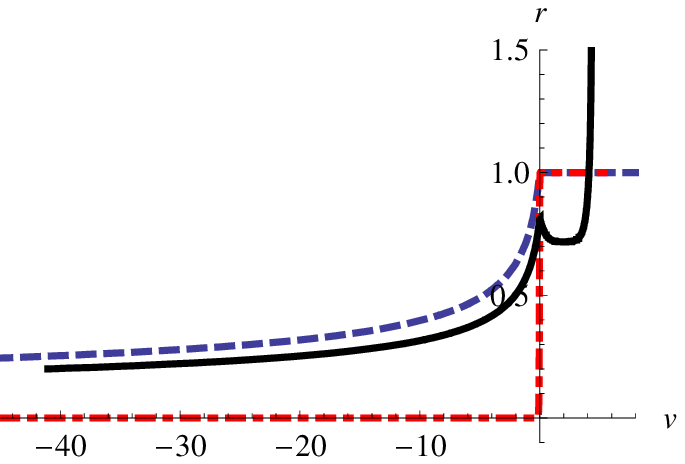}
\caption{\label{fig:geodesic} In both of the figures the dot-dashed curve (red) corresponds to the apparent horizon, the dashed curve (blue) to the event horizon, and the solid curve (black) to a generic geodesic contributing to the two point function. The left hand side figure corresponds to $z=2$ and the right hand side figure to $z=3$. The geodesic is seen to pass through both horizons, once through the event horizon and twice through the apparent horizon.}
\end{center}
\end{figure}

We can conclude that thermalization does not happen globally for the system at any finite time. Rather the thermalized region, as seen by the two point function, expands in time. This is the "horizon effect" observed in two point correlators in \cite{Calabrese:2006,Calabrese:2007rg}. The same effect has been observed before in holography in \cite{Albash:2010mv,Balasubramanian:2010ce,Balasubramanian:2011ur,Aparicio:2011zy} for the relativistic case $z=1$. The "horizon effect" seems more surprising for other values of $z$.

In Appendix~B we find the location of both the event horizon and the apparent horizon in the infalling shell spacetime with $z>1$. In an evolving geometry the apparent and event horizons can be very different and this is also seen here. As emphasized in \cite{AbajoArrastia:2010yt}, the two point correlation function probes regions of the spacetime that are inside both the event horizon and the apparent horizon. Indeed this seems necessary to reproduce the "horizon effect" known in quantum field theory. Figure~\ref{fig:geodesic} shows examples of geodesics, obtained numerically for different values of $z>1$, that pass through both horizons and also through the infalling shell into the $v<0$ region.

\begin{figure}[h]
\begin{center}
\includegraphics[scale=1.2]{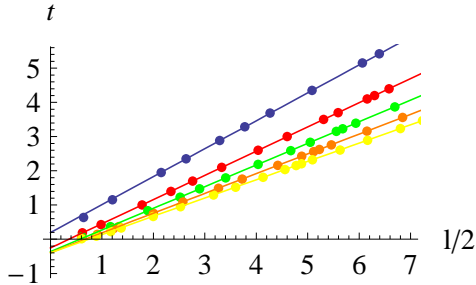}
\caption{\label{fig:thermtime} Lower bounds for thermalization times for different values of $z$. The data points are obtained by numerical integration of the integral in (\ref{eq:l}).  From top to bottom the different colored data points correspond to the values $z=1,2,3,4,5$. The lines with the corresponding colors are best fit lines with slopes given by (\ref{eq:velocity}).}
\end{center}
\end{figure}

Thermalization of the two point correlation function spreads with a finite velocity. Another way of saying this is that the thermalization time scale $t_{therm}(l)$ is an increasing function of $l$ with a finite slope. Here we define the thermalization time as the time when the two point function reaches the thermal value. First we can obtain a lower bound for the thermalization time $t_{therm}(l)$ in the case of a sharp quench. The correlator can be thermal only when the turning point of the geodesic $u_*$ is below the instantaneous position of the matter shell $u_0(t)$\footnote{So if the geodesic passes through the shell, the two point function will depend on time, and will thus not be thermal.}. We can obtain $u_0(t)$ by integrating (\ref{eq:nullgeod}). This leads to
\beq
t=\int_{0}^{u_0}\frac{du}{u^{-z+1}b(u)},\label{eq:t}
\eeq
where we use $b(u)=1-u^{2+z}$, since we are interested in the region outside the matter shell. The geodesic equations in the black brane backround can be integrated to
\beq
l=2\int_{0}^{u_*}\frac{du}{\sqrt{(\frac{u_*^2}{u^2}-1)b(u)}}.\label{eq:l}
\eeq
A lower bound for the thermalization time is when $u_0=u_*$ since for smaller times $t$ (or larger distances $l$) the geodesics necessarily pass through the shell and make the correlator time dependent. Note that this does not necessarily determine the real thermalization time as it can be that even at later times there is a geodesic which passes through the shell and has a shorter length than the one that probes the region outside the shell. The integral (\ref{eq:t}) can be seen to be an integral representation of the incomplete Beta function
\beq
t=\frac{1}{2+z}B(u_0^{z+2},\frac{z}{2+z},0),
\eeq
while we have not found a closed form for the integral in (\ref{eq:l}). Still we can find analytic expressions for both of the integrals in the limit as $t\rightarrow\infty$ and $l\rightarrow\infty$. This limit corresponds to integrating (\ref{eq:t}) close to the pole at $u_0\rightarrow1$, which leads to a logarithmic divergence. Similarly the integral in (\ref{eq:l}) diverges logarithmically as $u_*\rightarrow 1$. This leads to
\begin{align}
&t= -\frac{1}{2+z}\log(1-u_0)+\textrm{finite}\label{eq:falltime}
\\
&\frac{l}{2}= -\frac{1}{\sqrt{2(2+z)}}\log(1-u_*)+\textrm{finite}.
\end{align}
Identifying $u_0=u_*$ leads to a lower bound for the thermalization time
\beq
t=\frac{1}{\sqrt{1+z/2}}\frac{l}{2}+\textrm{finite},\label{eq:thermtimebound}
\eeq
for large $t$ and $l$. This gives us an upper bound for the velocity the thermalization can spread with
\beq
v=2\sqrt{1+z/2},\label{eq:velocity}
\eeq
for large $t$ and $l$. Transforming back to dimensionfull coordinates using (\ref{eq:scaling}) gives the dimensionfull velocity (recalling that in a theory with Lifshitz scaling symmetry velocity is indeed dimensionfull when $z\neq1$)
\beq
v=2\Big(\frac{4\pi T}{2+z}\Big)^{\frac{z-1}{z}}\sqrt{1+\frac{z}{2}}.\label{eq:dimvelocity}
\eeq
This is one of the main results of this section. Because of the finiteness of the bulk velocity of light (or causality in the bulk), there is an upper bound on how fast the two point function can thermalize, for all values of $z$. This upper bound depends on the state of the system after quench explicitly through the late time equilibrium temperature $T$ according to (\ref{eq:dimvelocity}). Also, this velocity is independent of the scaling dimension of the operator in the correlation function.\footnote{As long as the scaling dimension is sufficiently large for the geodesic approximation to hold.} A similar bound for general bulk spacetime dimension is calculated in Appendix A.

One should note that the fact that $l$ is linear function of $t$ in (\ref{eq:thermtimebound}) depends only on the presence of a first order zero in the metric function $b$ at the horizon. Thus, the "horizon" effect (that the thermalization of the two point function spreads in a cone with finite velocity) in a quantum quench is indeed directly related to the formation of a black hole horizon in the gravitational dual.

Even though the relation (\ref{eq:thermtimebound}) was derived in the limit of large $l$ and $t$ it can be seen to apply well for sufficiently small $l$ and $t$ as can be seen by evaluating the integrals (\ref{eq:t}) and (\ref{eq:l}) numerically. This comparison to the numerical evaluation is shown in Fig. \ref{fig:thermtime}

\begin{figure}[h]
\begin{center}
\includegraphics[scale=.25]{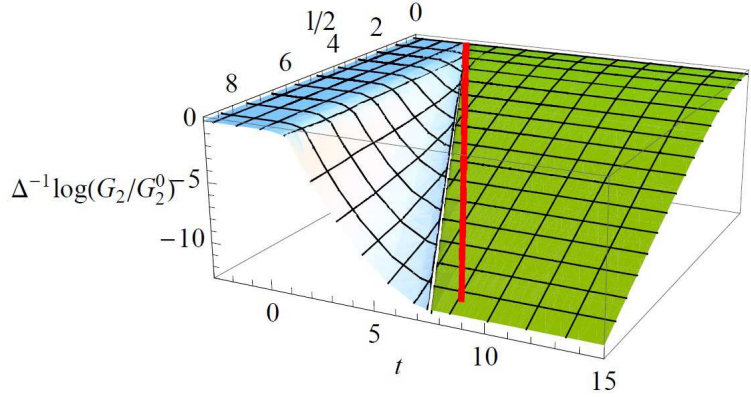}
\caption{\label{fig:quench1} Logarithm of the quench correlator for z=1, with the vacuum value subtracted. The red line corresponds to twice the speed of light. The blue surface corresponds to geodesics passing through the matter shell, while the green surface corresponds to geodesics probing the $u<1$ region.}
\end{center}
\end{figure}

\begin{figure}[h]
\begin{center}
\includegraphics[scale=.28]{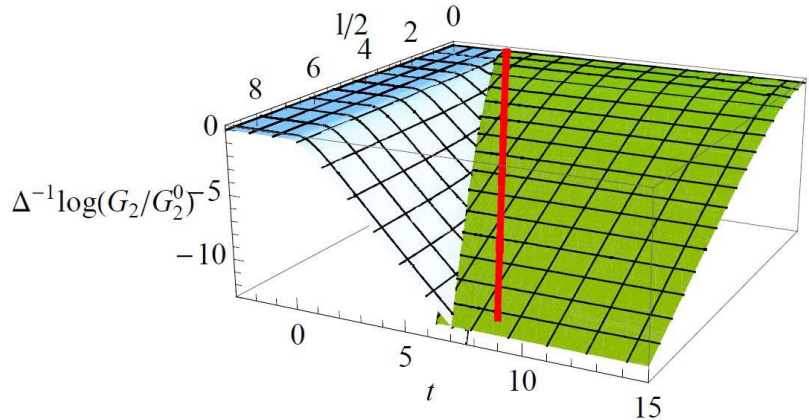}
\caption{\label{fig:quench2} Logarithm of the quench correlator for z=2, with the vacuum value subtracted. The red line is a reference line with slope $dl/dt=2$.}
\end{center}
\end{figure}

\begin{figure}[h]
\begin{center}
\includegraphics[scale=.25]{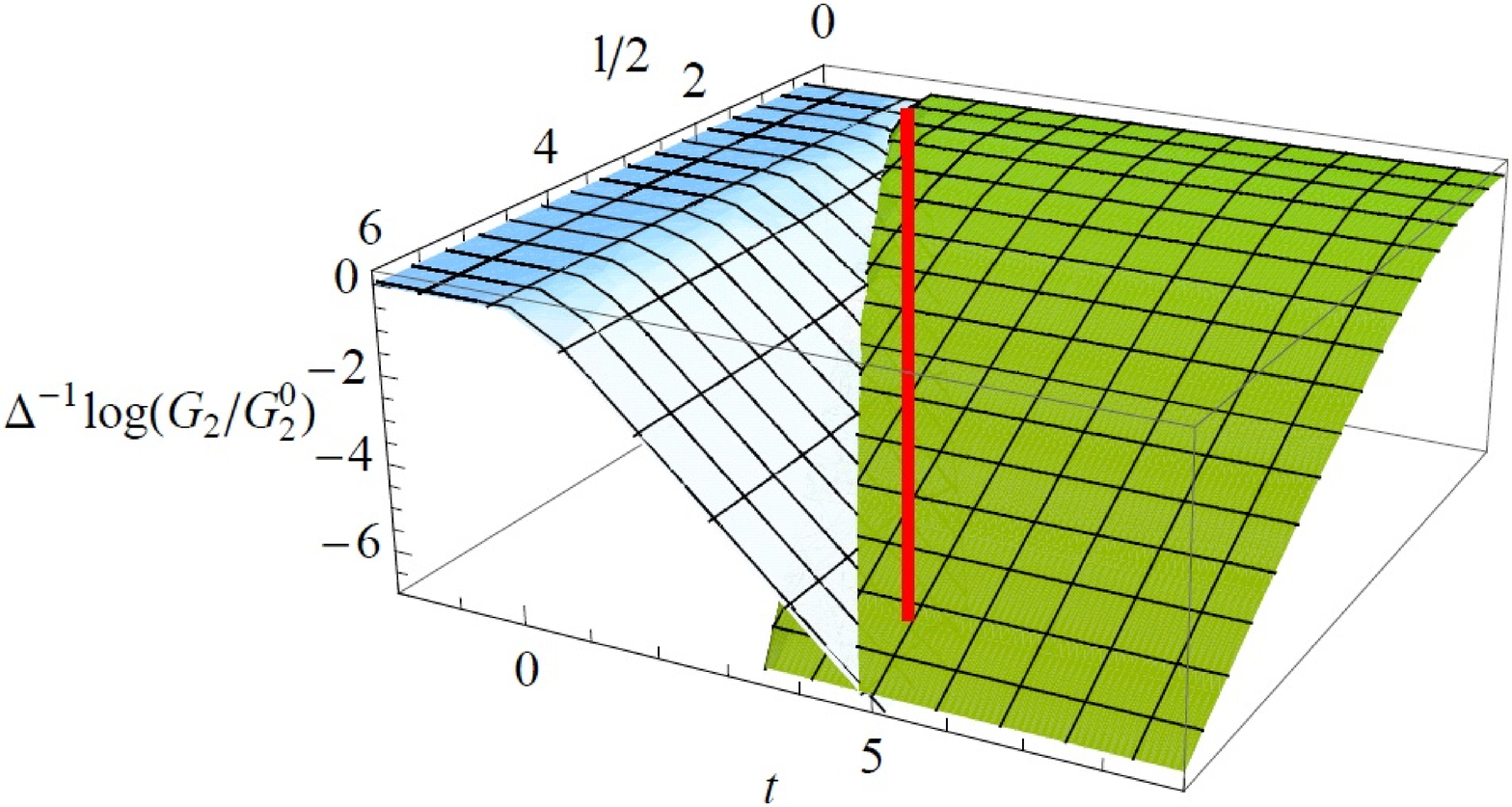}
\qquad
\includegraphics[scale=1.2]{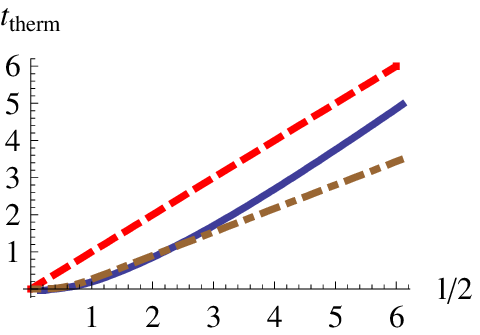}
\caption{\label{fig:quench3} a) Logarithm of the quench correlator for z=3 with the vacuum value subtracted. The red line is a reference line with slope $dl/dt=2$. The thermalization is seen to happen very suddenly as there are several geodesics contributing. b) The blue solid line is the real thermalization time extracted from the correlator, while the red dashed line is a reference line with slope 1 and the brown dot-dashed line is the lower bound for the thermalization time as obtained from numerical integration of (\ref{eq:t}) and (\ref{eq:l}).}
\end{center}
\end{figure}

Numerical results for the correlation functions for different values of $z$ are shown in Fig. \ref{fig:quench1}, Fig. \ref{fig:quench2} and Fig. \ref{fig:quench3}. For illustrational purposes we plot the logarithm of the correlator with the vacuum value substracted.

\begin{figure}[h]
\begin{center}
\includegraphics[scale=1.05 ]{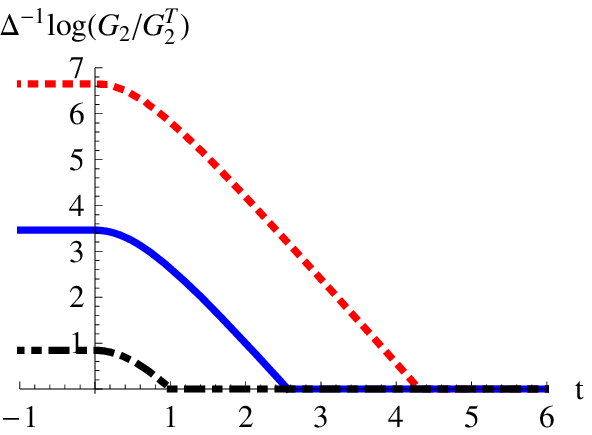}
\includegraphics[scale=1.1]{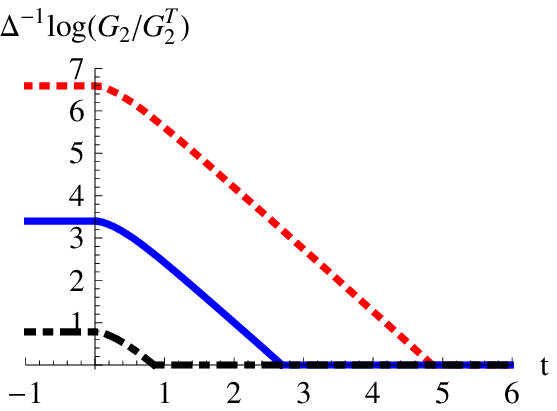}
\caption{\label{fig:quenchcorT} Logarithm of the two point correlator, with its thermal value subtracted. The different curves correspond to different values of $l$ as $l=4,8,12$ from bottom to top. The left figure is for $z=2$ while the right figure is for $z=3$.}
\end{center}
\end{figure}

To obtain the real thermalization time one has to take into account all the possible geodesics and pick the ones that have the lowest length to obtain the correlation function. For the ranges of $t$ and $l$ we have studied, the lower bound for the thermalization time agrees with the real thermalization time for $z=1$\footnote{There is nothing that guarantees this for larger and larger values of $l$.}. As can be seen from Fig. \ref{fig:quench2} and Fig. \ref{fig:quench3},\footnote{It should be noted that Fig. \ref{fig:quench3}b has been made by approximating the time dependent part of the correlator (the blue surface in Fig. \ref{fig:quench3}a ) as being independent of $l$. This is seen to be a good approximation at least for $l>4$. A similar approximation has been used to produce Fig. \ref{fig:quenchcorT} and in the next section Fig. \ref{fig:quenchent}b and Fig. \ref{fig:quenchentT}.} the real thermalization time for $z=2,3$ is indeed bigger than the lower bound as there are shorter geodesics available that pass through the shell. This is seen in the figures as the lengths of two branches of geodesics crossing. In the figure, the surface which is above, has a smaller geodesic length and corresponds to the real value of the correlation function.

It seems that there are two competing effects at work as $z$ is increased. On one hand, the upper bound for the thermalization velocity increases as $v\propto \sqrt{1+z/2}$, which can lead to faster thermalization as $z$ is increased. On the other hand it seems that for larger $z$, the real thermalization velocity due to competing geodesics is getting smaller than the upper bound leading to a slowing down of the velocity as can be seen from the right hand side of Fig. \ref{fig:quench3}.

From Fig. \ref{fig:quench2} and Fig. \ref{fig:quench3}a, we see that to a good approximation the ratio of the time dependent correlator and the vacuum correlator is independent of $l$ (for sufficiently large $l$) at a fixed time $t_0$. This tells us that outside the thermalized region the correlator can be approximated by
\beq
G_2(l,t_0)\approx\frac{Z(t_0)}{l^{2 \Delta}},
\eeq
where the wavefunction renormalization factor $Z$ is a decreasing function of the time. For the cases $z=2,3$, the correlation functions at fixed $l$ approach their thermal values linearly to a very good approximation as can be seen from Fig. \ref{fig:quenchcorT}. Thus we see that the time dependence of the wavefunction renormalization is to a good approximation given by
\beq
Z(t_0)\approx e^{-t_0\Delta/\tau},
\eeq
where $\tau$ is simply the inverse slope of the linear parts of the curves in Fig. \ref{fig:quenchcorT}. Also, Fig. \ref{fig:quenchcorT} shows that the time derivative of the correlator, as it reaches the thermal value, is discontinuous.

\newpage
\section{Entanglement entropy}

Another probe of correlations in the quench backround is provided by the entanglement entropy. A holographic formula for the entanglement entropy for a region $S$ in the boundary field theory has been suggested in \cite{Ryu:2006bv,Ryu:2006ef,Hubeny:2007xt} to be given by the area of a minimal surface that ends on a curve $C=\partial S$ at the AdS boundary. We will use the same prescription to calculate the entanglement entropy in the case of an asymptotically Lifshitz spacetime with $z\neq 1$. On the rest of this section we will simply assume that the above prescription for calculating the entanglement entropy extends to theories with Lifshitz scaling.

We will study the entanglement entropy of an infinite strip with a width $l$. The entanglement entropy is now obtained by minimizing the area functional
\beq
S_{ent}=\frac{1}{4G_N}\int d^2\sigma\sqrt{|\det\partial_{\alpha}X^{\mu}\partial_{\beta}X_{\mu}|}.
\eeq
For the infinite strip geometry we will choose to parametrize the surface with coordinates $\sigma^1=x,\sigma^2=y$ where the strip will be infinite in the $y$ direction and has a length $l$ in the $x$ direction. Due to translational symmetry of the strip in the $y$ direction, we can take an ansatz with $\partial_y X^{\mu}=0$. The coordinates of the surface are taken as
\beq
u=u(x),\quad v=v(x).
\eeq
Substituting this into the area functional gives
\beq
S_{ent}= \frac{1}{4G_N}\int dy\int dx u^{-1}\sqrt{-(v')^2u^{-2z}b(u,v)-2v'u'u^{-z-1}+u^{-2}}.
\eeq
It is convenient to define an entanglement entropy density as
\begin{align}
s_{ent}=\frac{4G_N S_{ent}}{\int dy}
&=\int dx u^{-1}\sqrt{-(v')^2u^{-2z}b(u,v)-2v'u'u^{-z-1}+u^{-2}}\nonumber
\\
&=\int dx L_A.\label{eq:entaction}
\end{align}
The fact that we multiply the entanglement entropy by $G_N$ can be interpreted in the dual field theory as dividing by the number of degrees of freedom in the dual field theory.\footnote{In the CFT case, by the central charge.} Again there is a conserved Hamiltonian following from the fact that the integrand in the area functional does not have explicit dependence on $x$
\beq
H_A=\frac{\partial L_A}{\partial u'}u'+\frac{\partial L_A}{\partial v'}v'-L_A=-\frac{1}{u^4 L_A}.\label{eq:hamiltonian}
\eeq
We will choose the $x$-coordinate in a way that the center of the strip is at $x=0$. Again the equations and boundary conditions are symmetric under $x\rightarrow -x$. Thus, the minimal length geodesic is again symmetric around $x=0$. Regularity requires us to set $u'(0)=v'(0)=0$. Denoting the turning point of the surface as $(u(0),v(0))=(u_*,v_*)$, we obtain the value of the hamiltonian (\ref{eq:hamiltonian}) as
\beq
H_A=-u_*^{-2}.
\eeq
This together with the Euler-Lagrange equation that follows from extremizing the area functional in (\ref{eq:entaction}) gives rise to the two independent equations
\begin{align}
uv''+4v'u'-2u^{z-1}+(z+1)u^{1-z}(v')^2-\frac{z}{2}m(v)u^3(v')^2&=0,\label{eq:are1}
\\
1-2u^{1-z}u'v'-u^{2-2z}b(u,v)(v')^2-\frac{u_*^4}{u^4}&=0.\label{eq:are2}
\end{align}

\subsection{Lifshitz vacuum}

In the Lifshitz vacuum $b=1$ and we can set $dv=-u^{z-1}du$. In this way (\ref{eq:are2}) becomes
\beq
1+(u')^2=\frac{u_*^4}{u^4},
\eeq
which can be integrated as
\beq
\frac{l}{2}=\int_{0}^{u_*}\frac{u^2du}{\sqrt{u_*^4-u^4}}=u_*\frac{\sqrt{\pi}\Gamma(3/4)}{\Gamma(1/4)}.
\eeq
The entanglement entropy density is now given by
\beq
s_{ent}=2\int_{0}^{l/2-\tilde{\epsilon}}dx\frac{u_*^2}{u^4}=2\int_{\epsilon}^{u_*}\frac{du u^{-2}}{\sqrt{1-(u/u_*)^4}},\label{eq:vacuument1}
\eeq
where $\tilde{\epsilon}$ is defined through $u(l/2-\tilde{\epsilon})=\epsilon$ and we used $du/dx=-\sqrt{u_*^4/u^4-1}$. Performing the integral in (\ref{eq:vacuument1}) we get
\beq
s_{ent}=\frac{2}{\epsilon}+\frac{1}{l}\frac{\pi\Gamma(-1/4)\Gamma(3/4)}{\Gamma(1/4)^2}.\label{eq:vacuument}
\eeq
This result is independent of $z$ and has the same value as in AdS spacetime \cite{Ryu:2006ef}, since again the spatial part of the Lifshitz vacuum metric is identical to that of AdS. The first term in (\ref{eq:vacuument}) is divergent as the boundary theory cutoff $\epsilon$ is sent to zero. Furthermore, since this term comes from the small $u$ behavior of the surface, it is independent of the state of the system. Meaning that it has the same value also in the thermal state and in the quench state. Thus, from now on we will substract the cutoff dependent term out of the entanglement entropy and define
\beq
\Delta s_{ent}=s_{ent}-\frac{2}{\epsilon}.
\eeq
The main physical physical point here is that
\beq
\Delta s_{ent}\propto -\frac{1}{l},\label{eq:vacent}
\eeq
which means that the entanglement entropy of a strip of width $l$ increases as $l$ is increased. Since $\Delta s_{ent}$ is the entanglement entropy per unit length in the $y$-direction, (\ref{eq:vacent}) is indeed consistent with dimensional analysis.

\subsection{Thermal equilibrium}

\begin{figure}[h]
\begin{center}
\includegraphics[scale=1.2]{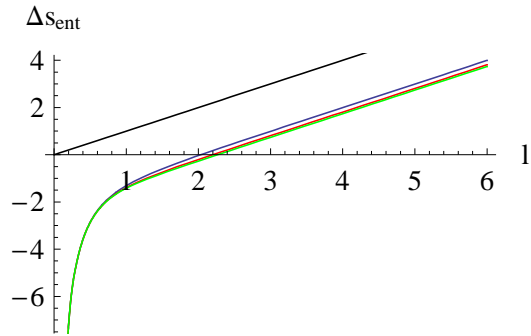}
\caption{\label{fig:thermalent} The finite part of the entanglement entropy density as a function of the width of the strip $l$, in the thermal state. The different curves correspond to the values $z=2,4,6$ from top to bottom. The straight line is a reference line with slope 1.}
\end{center}
\end{figure}

To study the entanglement entropy at finite temperature we set
\beq
b=1-u^{2+z},
\eeq
in equations (\ref{eq:are1}) and (\ref{eq:are2}). Next one can numerically integrate the equations. Results from a numerical integration are shown in Fig. \ref{fig:thermalent}.

The main physical points here are that the entanglement entropy for small distances behaves as in the vacuum
\beq
\Delta s_{ent}\propto -\frac{1}{l},\quad l \ll T^{-1/z},
\eeq
while for large distances the entanglement entropy seems to be a linear function of $l$
\beq
\Delta s_{ent}\propto l T^{2/z},\quad l \gg T^{-1/z},
\eeq
so that it seems extensive at large distances just as the usual thermal entropy.

\subsection{The quench}

To calculate the entanglement entropy in the quench state we should solve equations (\ref{eq:are1}) and (\ref{eq:are2}) for
\beq
b=1-m(v)u^{2+z},
\eeq
where $m(v)$ will again be chosen to be a hyperbolic tangent as in (\ref{eq:mass}). With the same arguments as we used in the case of the geodesic calculation in the previous section, also the entanglement entropy will thermalize only locally at any fixed value of time. Meaning that at large enough values of $l$ the entanglement entropy will be non-thermal and time dependent, no matter how large values of $t$ we look at. This again leads to a horizon effect similar to that in the two point function. Such a horizon effect has been observed before in the entanglement entropy in 1+1 dimensional CFTs in \cite{Calabrese:2005in} and later by using holography in \cite{AbajoArrastia:2010yt,Balasubramanian:2010ce,Albash:2010mv,Balasubramanian:2011ur}.

In fact the thermalization of the entanglement entropy will happen slower than that of the two point function in the holographic setup. We can see this simply by looking at the area functional in (\ref{eq:entaction}). Essentially this differs from the geodesic length functional simply by having a one extra power of $u^{-1}$ in the integrand. This means that in order to minimize the area functional, the turning point $u_*$ will want to be at a larger value of $u$ than that in the case of a geodesic. Thus, we see that the minimal surface will "hang" deeper in the bulk, and will pass through the matter shell with a smaller value of $l$. This means that the thermalization time for the entanglement entropy ought to be larger than that in the case of the two point function.

\begin{figure}[h]
\begin{center}
\includegraphics[scale=.22]{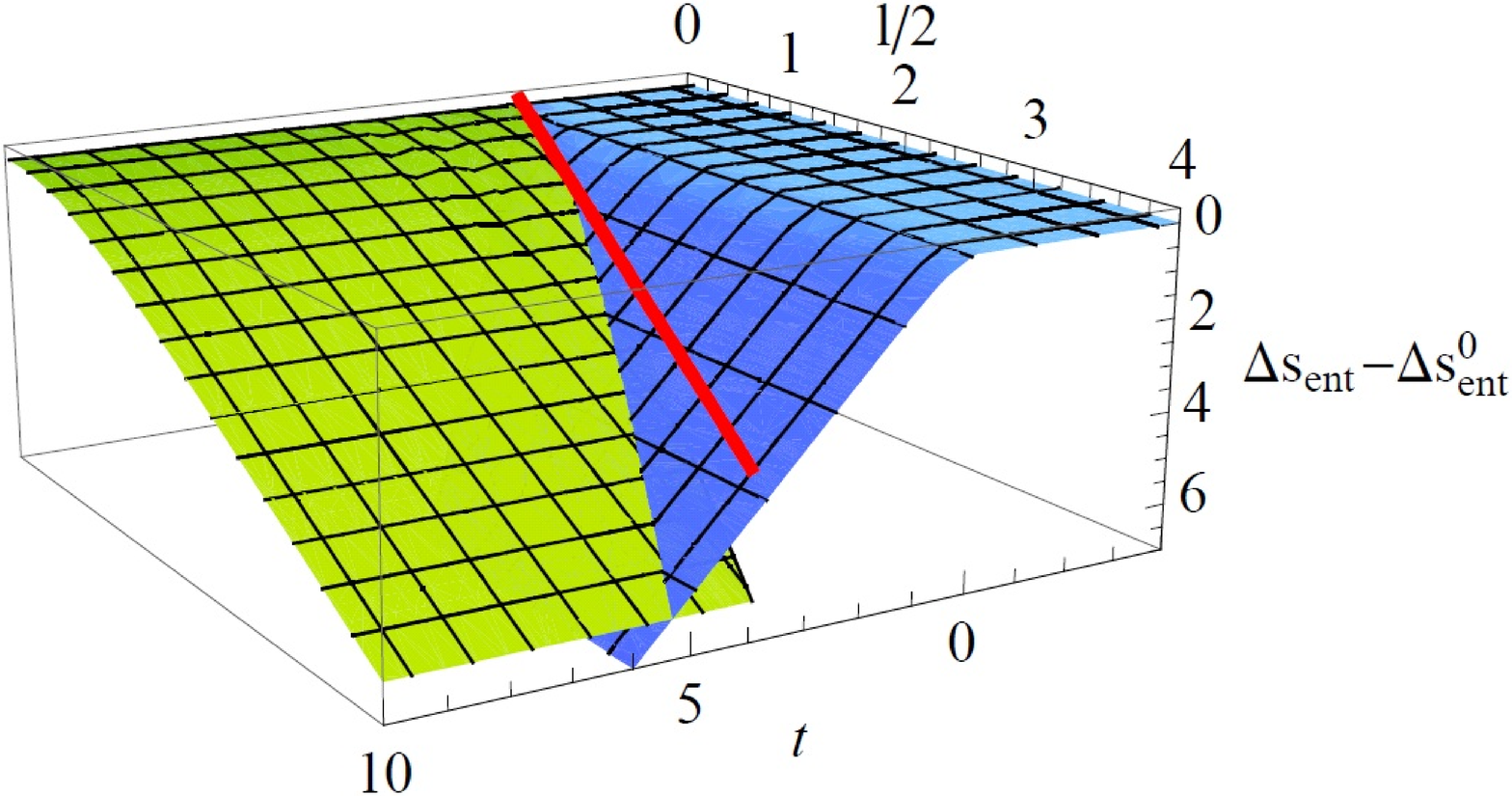}
\qquad
\includegraphics[scale=1.1]{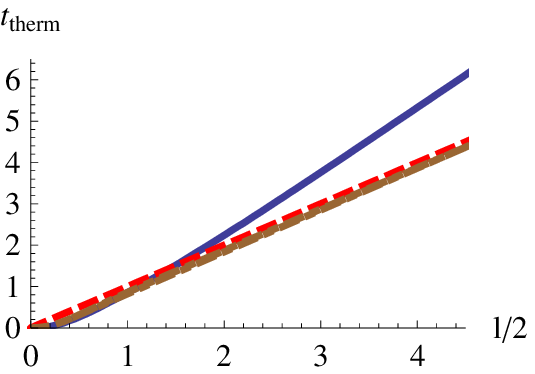}
\caption{\label{fig:quenchent} a) The entanglement entropy density in the quench state as a function of the width of the strip $l$ and the time $t$, for the case of $z=2$. The red line corresponds to the upper bound velocity, which in dimensionless coordinates is simply 1. b) The blue solid curve is the real thermalization time, while the brown dot-dashed curve is the lower bound for the thermalization time and the red dashed curve is a reference line with slope 1. As the figure shows, thermalization of the entanglement entropy really happens slower than the upper bound due to multiple surfaces contributing to the same values of $(l,t)$}.
\end{center}
\end{figure}

We can again make this intuition more precise in the case of a sharp quench (meaning $m(v)\propto \theta(v)$). The calculation is identical to that performed in the previous section except that now we the minimal surface equation integrates to
\beq
\frac{l}{2}=\int_{0}^{u_*}\frac{du}{\sqrt{(\frac{u_*^4}{u^4}-1)b(u)}}.
\eeq
This integral again diverges logarithmically as $u_*\rightarrow 1$ and we can extract the divergent part as
\beq
\frac{l}{2}=-\frac{1}{2\sqrt{2+z}}\log(1-u_*)+\textrm{finite}.
\eeq
Identifying $u_*$ with the position of the matter shell given by (\ref{eq:falltime}), we obtain an upper bound for the thermalization time (as the time when the minimal area surface starts to penetrate the matter shell) as
\beq
t=\frac{1}{\sqrt{\frac{1}{2}+\frac{z}{4}}}\frac{l}{2}+\textrm{finite}.
\eeq
Again this gives us an upper bound for the thermalization velocity (restoring the dimensionfull quantities from (\ref{eq:scaling}))
\beq
v=2\Big(\frac{4\pi T}{2+z}\Big)^{\frac{z-1}{z}}\sqrt{\frac{1}{2}+\frac{z}{4}}.\label{eq:dimvelocity2}
\eeq
The generalization of (\ref{eq:dimvelocity2}) to general bulk dimensionality is shown in Appendix A. This velocity is indeed smaller than that obtained for the thermalization of the two point function in (\ref{eq:dimvelocity}). Again we should emphasize that (\ref{eq:dimvelocity2}) is only an upper bound for the velocity the thermalization can propagate with. To obtain the real thermalization time one has to do the full numerical calculation of solving equations (\ref{eq:are1}) and (\ref{eq:are2}). The results from a numerical integration are shown in Fig. \ref{fig:quenchent} for the case of $z=2$. In this case we again see that the real thermalization velocity is smaller than the upper bound.

\begin{figure}[h]
\begin{center}
\includegraphics[scale=1.2]{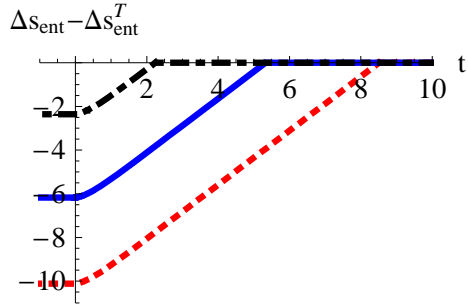}
\caption{\label{fig:quenchentT} Evolution of the entanglement entropy density in the quench state for $z=2$, with the thermal value subtracted. The different curves correspond to the values $l=4,8,12$.}.
\end{center}
\end{figure}

From Fig. \ref{fig:quenchentT} we see that the entanglement entropy increases to a very good approximation linearly with time. Also we see that the entanglement entropy reaches its thermal value very abruptly. Meaning that $\partial_tS_{ent}$ is discontinuous right when the entanglement entropy reaches a thermal value. A similar effect is seen for the case of $z=1$ in \cite{Balasubramanian:2010ce,Balasubramanian:2011ur,Albash:2010mv}.

\newpage

\section{Discussion}



In this paper we have studied quenches in theories with a scale invariance that is
anisotropic between space and time. The study of quenches in quantum
field theory has mainly concentrated on 1+1 dimensional systems near a
quantum critical point with relativistic scale invariance $z=1$, where one can use conformal
field theory methods \cite{Calabrese:2005in,Calabrese:2006,Calabrese:2007rg} to study the system. There one finds that after a quench the non-local observables such as two point correlation functions and entanglement entropy thermalize only locally at any fixed value of time. The size of the thermalized region is found to expand with twice the speed of light. This was called the "horizon" effect. More generally in "free" lattice systems it was found that the thermalized region expands with twice the group velocity of the slowest quasiparticle. This provides a nice intuitive picture of how the quench is followed by a release of a large number of energetic quasiparticles. As being released from sufficiently close to each other the quasiparticles are entangled. Subsequently these quasiparticles move with their corresponding group velocities (which in the case of a CFT is the speed of light). When a pair of quasiparticles reaches the points where the operators of the two point correlation function are located, one finds quantum correlations that are due to their initial entanglement. When the pair of quasiparticles has already passed through the two points, the correlation function is affected only by non-correlated quasiparticles, and it averages to a thermal value. When studying quantum quenches using holography a similar "horizon" effect was found in 1+1 as well as higher dimensional systems \cite{AbajoArrastia:2010yt,Balasubramanian:2010ce,Albash:2010mv,Balasubramanian:2011ur,Aparicio:2011zy}, albeit with some small differences.

The above results are only applicable to the case when the quantum critical point exhibits "relativistic" scaling $z=1$. In this paper we have studied quenches on quantum critical points with general values of $z$ using holography. We constructed analytic solutions corresponding to an infalling shell of massless and pressureless matter. These solutions are straightforward generalizations of the Vaidya solution \cite{Bonnor:1970zz} to an asymptotically Lifshitz spacetime. This solution provides a holographic model of a quench where energy is injected to the system at time $t=0$. Subsequently the system evolves into a thermal state, which in the gravitational description means that the matter shell forms a black hole.

In a quantum critical point with $z>1$, one expects the "quasiparticles" (if such a concept makes sense) to have a dispersion relation of the form $\omega\propto k^z$. This leads to a group velocity $v\propto k^{z-1}$. If one applies the heuristic picture of the "horizon" effect to these "quasiparticles", one concludes that thermalization of non-local observables never happens as the group velocity of the slowest modes is arbitrarily close to zero. On the other hand, if the thermalization is somehow driven by the high energy modes one might think that thermalization happens immediately as $v$ is not bounded from above for large $k$.

Neither of these possibilities is what we find in the holographic model. What we find is that there is a "horizon" effect which proceeds at a finite velocity $v\propto T^{(z-1)/z}$, where $T$ is the temperature of the equilibrium state reached. The peculiar temperature dependence of the propagation velocity of the thermalization "horizon" follows simply from dimensional analysis. It also suggests that the thermalization is mainly mediated by "quasiparticles" with average energy $T$ as this can lead to a velocity $v\propto T^{(z-1)/z}$. This is of course natural since after the quench, the average energy of a "quasiparticle" should indeed be given by $T$. We do not find any time dependence in the correlation function or the entanglement entropy after the thermalization time is passed. So it seems that there are no "small momentum quasiparticles" that could destroy the thermalization, as they would lead to late time oscillations in the correlators \cite{Calabrese:2007rg}. A possible explanation is that any "quasiparticles" with small momenta will be excited to have momenta of the order $k=T^{1/z}$ by scatterings with the surrounding medium, as we are dealing with theories which are strongly interacting. Then because the "quasiparticle" group velocity is a non-trivial function of the momentum (this is different from the case $z=1$ where the velocity is a constant, the speed of light) the "quasiparticles" can speed up to $v\propto T^{(z-1)/z}$ as given by the average energy in the system. Similarly, high energy "quasiparticles" can slow down to velocities $v\propto T^{(z-1)/z}$ through scatterings.

We also found upper bounds for the velocity of the thermalization "horizon" analytically. These indeed show that thermalization must happen at a finite velocity even though there is no reason for this from causality in the dual field theory for the case of $z>1$. The velocity bounds, as well as the real thermalization velocities, are different for different observables. We found that generally entanglement entropy is the observable that thermalizes last. The ratio between the thermalization velocities of two point functions and entanglement entropy was found to be order 1.

Even though the different observables thermalized with different velocities, there was some universality in the two point functions because the velocity at which they thermalized was seen to be independent of the scaling dimensions of the operators.\footnote{This may be an artifact of the geodesic approximation.}

\section{Acknowledgements}

This work was supported in part by the Icelandic Research Fund and
by the University of Iceland Research Fund. E.K-V. has been supported in part by the Academy of Finland grant number 1127482. We would like to thank many colleagues for discussions on thermalization and holography, and
Niels Obers for a discussion on Lifshitz black holes. We would like to thank  A. Bernamonti and F. Galli for useful comments on the manuscript. V.K. would like to thank Helsinki Institute of Physics,
and E.K-V. and L.T. the Imperial College London for hospitality during the early stage of this work.

\newpage

\appendix
\section{A higher dimensional example}

In the bulk of the text we were dealing with the case where the bulk was $3+1$ dimensional. The results for higher dimensional cases are qualitatively similar. In this appendix we consider $d$ dimensional bulk, and present numerical results for the case $d=5$ and $z=2$.

\begin{figure}[h]
\begin{center}
\includegraphics[scale=.22]{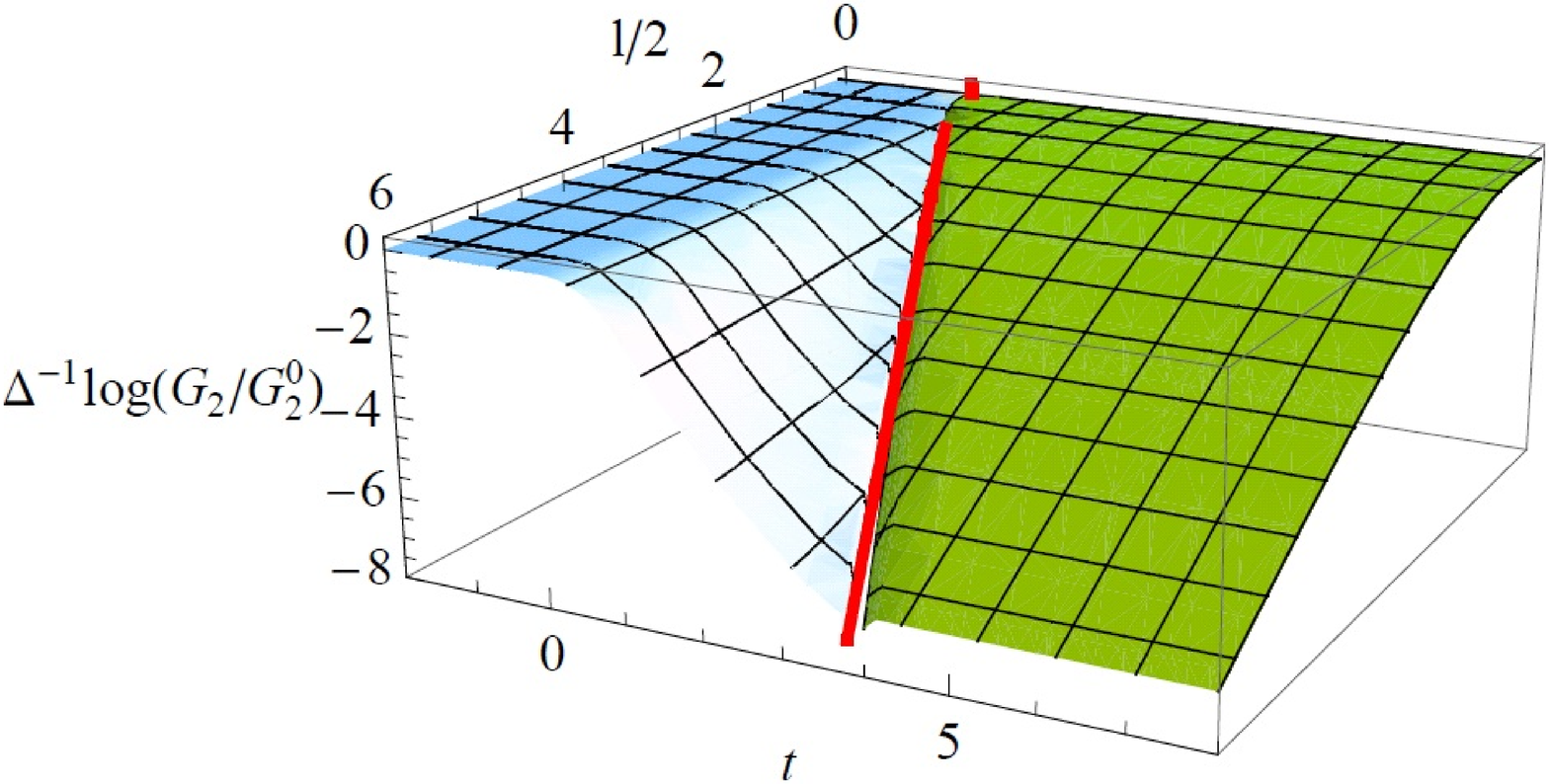}
\quad
\includegraphics[scale=.22]{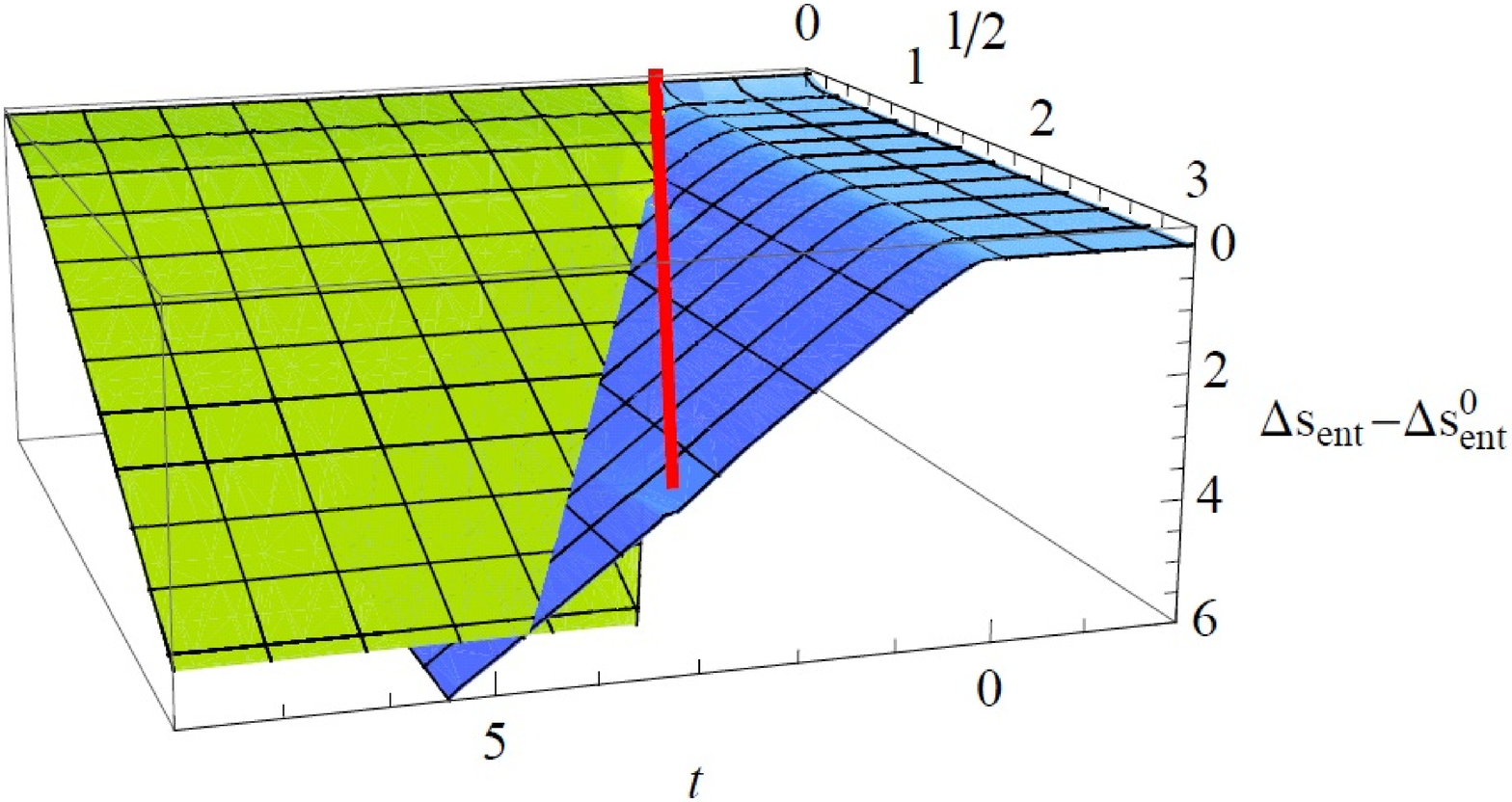}
\caption{\label{fig:quenchcord} a) The logarithm of the two point correlation function with vacuum value subtracted, for $z=2$ and 3+1 dimensional dual field theory. The red curve corresponds to the upper bound speed $v=\sqrt{10}$ from (\ref{eq:upperspeed}). b) The entanglement entropy density for the same $d$ and $z$. The red curve corresponds to the upper bound speed $v=\sqrt{10/3}$ from (\ref{eq:upperspeedent}).}
\end{center}
\end{figure}

The geodesic equations (\ref{eq:geo1}) and (\ref{eq:geo2}) depend on the bulk dimensionality only through the metric factor
\beq
b(u,v)=1-m(v)u^{d-2+z}.
\eeq
Thus, they generalize immediately to general values of $d$. By considering the case of a thin shell, we can again calculate analytically a lower bound for the thermalization time at large values of $l$. The integrals (\ref{eq:t}) and (\ref{eq:l}) apply to the higher dimensional case with the difference that the metric factor is $b=1-u^{d-2+z}$. With this difference, the logarithmically divergent parts are now
\begin{align}
t&=-\frac{1}{d+z-2}\log(1-u_*)+\textrm{finite},\label{eq:t2}
\\
\frac{l}{2}&=-\frac{1}{\sqrt{2(d+z-2)}}\log(1-u_*)+\textrm{finite}.
\end{align}
transforming back to dimensionfull quantities we get an upper bound velocity (at large $l$) with which the thermalization can spread
as
\beq
v=2\Big(\frac{4\pi T}{2+z}\Big)^{\frac{z-1}{z}}\sqrt{\frac{d+z-2}{2}}.\label{eq:upperspeed}
\eeq
As an example for the full correlation function the case $d=5$ and $z=2$ is displayed in Fig. \ref{fig:quenchcord}. Thermalization is in this case seen to spread with the upper bound velocity, which in this case is (in dimensionless coordinates) $v=\sqrt{10}$.

The entanglement entropy is now obtained by minimizing
\beq
S_{ent}=\frac{1}{4G_N}\int d^{d-2}\sigma\sqrt{|\det \partial_{\alpha}X^{\mu}\partial_{\beta}X_{\mu}|}.
\eeq
Here we calculate the entanglement entropy for a "belt" region. In this way we will choose the coordinates parametrizing the hypersurface as $\sigma^1=x, \sigma^2=y,...,\sigma^{d-2}=w$. The interior of the "belt" corresponds to $x\in (-l/2,l/2)$ and the other coordinates range from $-\infty$ to $+\infty$. In this way the entanglement entropy density functional to minimize becomes
\beq
s_{ent}=\frac{4G_N S_{ent}}{\int dy...dw}=\int dx u^{3-d}\sqrt{-(v')^2u^{-2z}b(u,v)-2v'u'u^{-z-1}+u^{-2}}.\label{eq:ent2}
\eeq
Minimizing (\ref{eq:ent2}) leads to the equations
\begin{align}
&uv''+2(d-2)v'u'-(d-2)u^{z-1}+(z+d-3)u^{1-z}(v')^2\nonumber
\\
&+\frac{1}{2}(4-z-d)u^{d-1}m(v)(v')^2=0,\label{eq:are1d}
\\
&1-2u^{1-z}u'v'-u^{2-2z}b(u,v)(v')^2-\frac{u_*^{2d-4}}{u^{2d-4}}=0.\label{eq:are2d}
\end{align}
Again we can analytically calculate a lower bound for the thermalization time. This time the minimal surface equations lead to an integral
\beq
\frac{l}{2}=\int_0^{u_*}\frac{du}{\sqrt{(\frac{u_*^{2d-4}}{u^{2d-4}}-1)b(u)}}=-\frac{1}{\sqrt{2(d-2)(z+d-2)}}\log(1-u_*)+\textrm{finite},
\eeq
which together with (\ref{eq:t2}) leads to an upper bound velocity (in dimensionfull coordinates)
\beq
v=2\Big(\frac{4\pi T}{2+z}\Big)^{\frac{z-1}{z}}\sqrt{\frac{d+z-2}{2(d-2)}}\label{eq:upperspeedent}
\eeq
Curiously, the ratio of the upper bound velocities for the two point function and the entanglement entropy is independent of $z$ and is given by $\sqrt{d-2}$. The entanglement entropy for $z=2$ and $d=5$ as a function of $l$ and $t$ is shown in Fig. \ref{fig:quenchcord}b. In this case thermalization happens slower than the upper bound velocity.

\section{Apparent horizon and event horizon}

In this appendix we locate the apparent horizon and the event horizon for the infalling shell metric
\beq
ds^2=-r^{2z}b(r,v)dv^2+2r^{z-1}dvdr+r^2d\textbf{x}^2,
\eeq
where we take the number of spacetime dimensions to be $d=4$. The calculation follows closely the corresponding calculation for $z=1$ in \cite{AbajoArrastia:2010yt}.

Our spacetime can be foliated by the spacelike surfaces $r=$const and $v=$const and then we can define two null vectors that are orthogonal to the surfaces
\beq
N_{in}=-r^{1-z}\partial_r,\quad N_{out}=\partial_v+\frac{1}{2}r^{z+1}b(r,v)\partial_r,
\eeq
where the normalization is chosen in a way that $N_{in}\cdot N_{out}=-1$. The expansion around the null directions are given by $\theta=\tilde{g}^{\mu\nu}\nabla_{\mu}N_{\nu}$, where $\tilde{g}^{\mu\nu}$ is the induced metric on the surface. A straightforward calculation gives
\beq
\theta_{in}=-2 r^{-z},\quad\theta_{out}=r^{z}b(r,v).
\eeq
Thus, the inward null geodesics always converge, while the outward geodesics converge when $b(r,v)<0$. The apparent horizon is in this case given by the locus where $b(r,v)=0$. For the uncharged black hole solutions, this gives the position of the apparent horizon as
\beq
r(v)=m(v)^{\frac{1}{2+z}}.
\eeq
Now we calculate the position of the event horizon. Since in our case the mass function $m(v)$ approaches a constant $m$ as $v\rightarrow\infty$, the event horizon is simply given by the outgoing null geodesic which approaches the position of the apparent horizon as $v\rightarrow\infty$. Outgoing null geodesics satisfy the first order differential equation
\beq
\frac{dr}{dv}=\frac{1}{2}r^{z+1}b(r,v).
\eeq
For a given mass function the solution to this equation, with the boundary condition $\lim_{v\rightarrow\infty}r(v)=m^{1/(2+z)}$, specifies the event horizon uniquely. The event horizon and the apparent horizon are shown in Fig. \ref{fig:geodesic}, for the choice of mass function in (\ref{eq:mass}).


\newpage
\begin{thebibliography}{99}

\bibitem{bloch}
  I.~Bloch, J.~Dalibard, W.~Zwerger
  ``Manybody Physics with Ultracold Gases,''
  Rev.\ Mod.\ Phys.\ {\bf 80}, (2008) 885,
  [arXiv:0704.3011 [cond-mat]].

\bibitem{greiner1}
  M.~Greiner, O.~Mandel, T.~Esslinger, T.~W.~Hänsch, I.~Bloch,
  ``Quantum phase transition from a superfluid to a Mott insulator in a gas of ultracold atoms,''
  Nature {\bf 415}, (2002) 39-44.

\bibitem{greiner2}
  M.~Greiner, O.~Mandel, T.~W.~Hänsch, I.~Bloch,
  ``Collapse and revival of the matter wave field of a Bose–Einstein condensate,''
  Nature {\bf 419}, (2002) 51-54.

\bibitem{Calabrese:2005in}
  P.~Calabrese, J.~L.~Cardy,
  ``Evolution of entanglement entropy in one-dimensional systems,''
  J.\ Stat.\ Mech.\  {\bf 0504 } (2005)  P04010.
  [arXiv:0503393 [cond-mat]].

\bibitem{Calabrese:2006}
  P.~Calabrese, J.~Cardy,
  ``Time-dependence of correlation functions following a quantum quench,''
  Phys.\ Rev.\ Lett\ {\bf 96 } (2006) 136801.
  [arXiv:0601225 [cond-mat]].

\bibitem{DeGrandi}
  C.~De~Grandi, V.~Gritsev, A.~Polkovnikov,
  ``Quench Dynamics Near a Quantum Critical Point,''
  Phys.\ Rev.\ {\bf B81}, (2010) 012303,
  [arXiv:0909.5181v3 [cond-mat.stat-mech]].

\bibitem{Calabrese:2007rg}
  P.~Calabrese, J.~Cardy,
  ``Quantum Quenches in Extended Systems,''
  J.\ Stat.\ Mech.\  {\bf 0706 } (2007)  P06008.
  [arXiv:0704.1880 [cond-mat.stat-mech]].

\bibitem{jacek}
  J.~Dziarmaga,
  ``Dynamics of a Quantum Phase Transition and Relaxation to a Steady State,''
  [arXiv:0912.4034 [cond-mat]].


\bibitem{AbajoArrastia:2010yt}
  J.~Abajo-Arrastia, J.~Aparicio, E.~Lopez,
  ``Holographic Evolution of Entanglement Entropy,''
  JHEP {\bf 1011 } (2010)  149.
  [arXiv:1006.4090 [hep-th]].

\bibitem{Balasubramanian:2010ce}
  V.~Balasubramanian, A.~Bernamonti, J.~de Boer, N.~Copland, B.~Craps, E.~Keski-Vakkuri, B.~Muller, A.~Schafer {\it et al.},
  ``Thermalization of Strongly Coupled Field Theories,''
  Phys.\ Rev.\ Lett.\  {\bf 106 } (2011)  191601.
  [arXiv:1012.4753 [hep-th]].

\bibitem{Albash:2010mv}
  T.~Albash, C.~V.~Johnson,
  ``Evolution of Holographic Entanglement Entropy after Thermal and Electromagnetic Quenches,''
  New J.\ Phys.\  {\bf 13 } (2011)  045017.
  [arXiv:1008.3027 [hep-th]].

\bibitem{Balasubramanian:2011ur}
  V.~Balasubramanian, A.~Bernamonti, J.~de Boer, N.~Copland, B.~Craps, E.~Keski-Vakkuri, B.~Muller, A.~Schafer {\it et al.},
  ``Holographic Thermalization,''
  Phys.\ Rev.\  {\bf D84 } (2011)  026010.
  [arXiv:1103.2683 [hep-th]].

\bibitem{Aparicio:2011zy}
  J.~Aparicio, E.~Lopez,
  ``Evolution of Two-Point Functions from Holography,''
  [arXiv:1109.3571 [hep-th]].

\bibitem{Balasubramanian:2011at}
  V.~Balasubramanian, A.~Bernamonti, N.~Copland, B.~Craps, F.~Galli,
  ``Thermalization of mutual and tripartite information in strongly coupled two dimensional CFTs,''
  [arXiv:1110.0488 [hep-th]].

\bibitem{Allais:2011ys}
  A.~Allais, E.~Tonni,
  ``Holographic evolution of the mutual information,''
  [arXiv:1110.1607 [hep-th]].

\bibitem{Das:2010yw}
  S.~R.~Das, T.~Nishioka, T.~Takayanagi,
  ``Probe Branes, Time-dependent Couplings and Thermalization in AdS/CFT,''
  JHEP {\bf 1007}, 071 (2010).
  [arXiv:1005.3348 [hep-th]].

\bibitem{Basu:2011ft}
  P.~Basu and S.~R.~Das,
  ``Quantum Quench across a Holographic Critical Point,''
  arXiv:1109.3909 [hep-th].

\bibitem{Kachru:2008yh}
  S.~Kachru, X.~Liu and M.~Mulligan,
  ``Gravity Duals of Lifshitz-like Fixed Points,''
  Phys.\ Rev.\  D {\bf 78} (2008) 106005
  [arXiv:0808.1725 [hep-th]].

\bibitem{Koroteev:2007yp}
  P.~Koroteev, M.~Libanov,
  ``On Existence of Self-Tuning Solutions in Static Braneworlds without Singularities,''
  JHEP {\bf 0802}, 104 (2008).
  [arXiv:0712.1136 [hep-th]].


\bibitem{Hartnoll:2011fn}
  S.~A.~Hartnoll,
  ``Horizons, holography and condensed matter,''
  [arXiv:1106.4324 [hep-th]].






\bibitem{Gubser:2009cg}
  S.~S.~Gubser, A.~Nellore,
  ``Ground states of holographic superconductors,''
  Phys.\ Rev.\  {\bf D80 } (2009)  105007.
  [arXiv:0908.1972 [hep-th]].

\bibitem{Hartnoll:2009ns}
  S.~A.~Hartnoll, J.~Polchinski, E.~Silverstein, D.~Tong,
  ``Towards strange metallic holography,''
  JHEP {\bf 1004}, 120 (2010).
  [arXiv:0912.1061 [hep-th]].

\bibitem{Hartnoll:2010gu}
  S.~A.~Hartnoll, A.~Tavanfar,
  ``Electron stars for holographic metallic criticality,''
  Phys.\ Rev.\  {\bf D83 } (2011)  046003.
  [arXiv:1008.2828 [hep-th]].

\bibitem{Vaidya:1951zz}
  P.~Vaidya,
  ``The Gravitational Field of a Radiating Star,''
  Proc.\ Indian Acad.\ Sci.\  {\bf A33 } (1951)  264.

\bibitem{Bonnor:1970zz}
  W.~B.~Bonnor, P.~C.~Vaidya,
  ``Spherically symmetric radiation of charge in Einstein-Maxwell theory,''
  Gen.\ Rel.\ Grav.\  {\bf 1 } (1970)  127-130.


\bibitem{Bhattacharyya:2009uu}
  S.~Bhattacharyya, S.~Minwalla,
  ``Weak Field Black Hole Formation in Asymptotically AdS Spacetimes,''
  JHEP {\bf 0909 } (2009)  034.
  [arXiv:0904.0464 [hep-th]].

\bibitem{Ebrahim:2010ra}
  H.~Ebrahim, M.~Headrick,
  ``Instantaneous Thermalization in Holographic Plasmas,''
  [arXiv:1010.5443 [hep-th]].

\bibitem{Taylor:2008tg}
  M.~Taylor,
  ``Non-relativistic holography,''
  [arXiv:0812.0530 [hep-th]].

\bibitem{Tarrio:2011de}
  J.~Tarrio, S.~Vandoren,
  ``Black holes and black branes in Lifshitz spacetimes,''
  JHEP {\bf 1109}, 017 (2011).
  [arXiv:1105.6335 [hep-th]].


\bibitem{Ryu:2006bv}
  S.~Ryu, T.~Takayanagi,
  ``Holographic derivation of entanglement entropy from AdS/CFT,''
  Phys.\ Rev.\ Lett.\  {\bf 96 } (2006)  181602.
  [hep-th/0603001].

\bibitem{Ryu:2006ef}
  S.~Ryu, T.~Takayanagi,
  ``Aspects of Holographic Entanglement Entropy,''
  JHEP {\bf 0608 } (2006)  045.
  [hep-th/0605073].

\bibitem{Hubeny:2007xt}
  V.~E.~Hubeny, M.~Rangamani and T.~Takayanagi,
  ``A Covariant holographic entanglement entropy proposal,''
  JHEP {\bf 0707} (2007) 062
  [arXiv:0705.0016 [hep-th]].



\bibitem{Solodukhin:2009sk}
  S.~N.~Solodukhin,
  ``Entanglement Entropy in Non-Relativistic Field Theories,''
  JHEP {\bf 1004}, 101 (2010)
  [arXiv:0909.0277 [hep-th]].

\bibitem{Nesterov:2010yi}
  D.~Nesterov and S.~N.~Solodukhin,
  ``Gravitational effective action and entanglement entropy in UV modified
  theories with and without Lorentz symmetry,''
  Nucl.\ Phys.\  B {\bf 842}, 141 (2011)
  [arXiv:1007.1246 [hep-th]].

\bibitem{deBoer:2011wk}
  J.~de Boer, M.~Kulaxizi and A.~Parnachev,
  ``Holographic Entanglement Entropy in Lovelock Gravities,''
  JHEP {\bf 1107}, 109 (2011)
  [arXiv:1101.5781 [hep-th]].

\bibitem{Balasubramanian:1999zv}
  V.~Balasubramanian, S.~F.~Ross,
  ``Holographic particle detection,''
  Phys.\ Rev.\  {\bf D61 } (2000)  044007.
  [hep-th/9906226].

\bibitem{Banks:1998dd}
  T.~Banks, M.~R.~Douglas, G.~T.~Horowitz, E.~J.~Martinec,
  ``AdS dynamics from conformal field theory,''
  [hep-th/9808016].


\end{thebibliography}
\end{document}